\documentclass{article}

\usepackage[latin1]{inputenc} 
\usepackage[T1]{fontenc} 

\usepackage{amsmath}
\usepackage{amssymb}
\usepackage{a4wide}
\usepackage{graphicx} 
\usepackage{color}
\usepackage{epic}

\newcommand{\dps}{\displaystyle}

\begin{document}

\title{A reduced model for shock and detonation waves. I. The inert case.}
\author{Gabriel STOLTZ$^{1,2}$ \\
\footnotesize{1: CERMICS, Ecole Nationale des
  Ponts et Chauss\'ees (ParisTech),}\\ \footnotesize{6 \& 8 Av. Pascal,
  77455 Champs-sur-Marne, France.} \\
\footnotesize{2: CEA/DAM Ile-de-France,
BP 12, 91680 Bruy\`eres-le-Ch\^atel, France.}\\
\footnotesize{ stoltz@cermics.enpc.fr} 
}

\maketitle

\begin{abstract}
We present a model of mesoparticles, very much in the Dissipative Particle Dynamics spirit, in which a molecule is replaced by a particle with an internal thermodynamic degree of freedom (temperature or energy). The model is shown to give quantitavely accurate results for the simulation of shock waves in a crystalline polymer, and opens the way to a reduced model of detonation waves. 
\end{abstract}

\section{Introduction}

Multimillion atom simulations are nowadays common in molecular dynamics (MD) studies. However, the time and space scales numerically tractable are still far from being macroscopic, so that reduced models are of primary interest when multiscale phenomena are considered.
In particular, the simulation of shock waves is a challenging task, involving very small time and space scales and large energies near the shock front, and much larger time and space scales and lower energies for the relaxation of the shocked materials, including the evolution of dislocations loops for example. 

Some reduced models for shock waves were proposed, for polycrystalin materials~\cite{Holian03}, crystalin materials with a projection of the dynamics onto a one-dimensional atom chain~\cite{stoltz05}, or resorting to mesoparticles with internal degrees of freedom~\cite{SH04}. This last approach seems to be the most promising and the most general one, and consists in replacing a complex molecule by a single particle. The introduction of an internal degree of freedom describing in a mean way the behavior of several degrees of freedom is reminiscent from Dissipative Particle Dynamics (DPD) models, which aim at describing complex fluids through some mesodynamics with some additional variables. 

DPD models, introduced in~\cite{HK92}, have been put on firm thermodynamics ground in~\cite{EW95}. Some derivations from MD where proposed in a simplified case in~\cite{Espanol96}, the more convincing general derivation being at the moment~\cite{FCdF00}. These studies motivate the modelling of the mean action of the projected degrees of freedom through some dissipative forces (depending on the relative velocities of the particles, so that the global momentum is conserved), balanced by some random forces.
Ergodicity of the dynamics can be shown in some simplified cases~\cite{SY06}. Therefore, DPD dynamics are well established and motivated reduced models. 

Coarser models such as SPH (Smoothed particle hydrodynamics)~\cite{Lucy77,Monaghan92} are routinely used to simulate shock waves at the hydrodynamic level, and can also be formulated in a DPD framework (the so-called Smoothed dissipative particle dynamics~\cite{ER03}). However, these models require the knowledge of an equation of state $E_{\rm int} = E_{\rm int}(S,P)$ giving the internal energy as a function of entropy and pressure, for instance. Therefore, SPH-like models cannot be considered when the coarse-grained model is still at the microscopic level. 

It is also not possible to resort directly to the classical DPD models to simulate shock waves. 
Indeed, the dissipative and random forces arising in DPD are linked through some fluctuation-dissipation relation, using a local temperature. But when a shock wave passes, energy is transfered to the material, and the local temperature changes.
Therefore, it is necessary to consider DPD models where the fluctuation-dissipation relation is not fixed {\it a priori}, but evolves depending on the physical events that have happened. DPD with conserved energy~\cite{AM97,Espanol97} are such models. 
Let us emphasize at this point that keeping thermal fluctuations in the microscopic models is of paramount importance to obtain the right relaxation profiles behind the shock front~\cite{Holian95,stoltz05}.

We present here a dynamics strongly inspired by those models, and show that it provides an interesting mesoscopic model for the simulation of shock waves. To our knowledge, this is the first study on shock waves using stochastic dynamics of this kind. 
It also opens the way for an extension to detonation waves~\cite{MSS}, where exothermic chemical reactions are triggered as the shock passes, with the shock sustained and enhanced through the energy released. 

The paper is organized as follows. We first recall the usual DPD equations, and
propose our mesoscopic dynamics. We then
discuss the numerical implementation of the dynamics and present some simulation results.

\section{Dissipative Particle Dynamics with conserved energy}
\label{DPD_models}

All atom simulations are performed resorting to Newton's equations of motion. The corresponding microscopic systems are deterministic, Galilean invariant, and have some invariants, such as the total energy. While stochastic models are natural models to describe systems with reduced dynamics (since the information lost by the averaging process is modelled by some random process), it is however not clear that such a stochastic model can reproduce, even in a mean way, a deterministic dynamics with invariants.

It turns out however that DPD models are stochastic dynamics which are
Galilean invariant and preserve total momentum. Some refinements were also proposed in order
to conserve the total energy of the system, a model called 'DPD with conserved energy' (DPDE~\cite{AM97,Espanol97}). 

We consider a system of $N$ particles in a space of dimension $d$, described by their positions $(q_1,\dots,q_N)$ and momenta $(p_1,\dots,p_N)$, with associated mass matrix $M = {\rm Diag}(m_1,\dots,m_N)$, interacting through a potential ${\cal V}$. We assume for simplicity that the interactions between the particles are pairwise and depend only on the relative distances, so that ${\cal V}(q) = \sum_{i < j} V(|q_i-q_j|)$.
Denoting by $\bar{T}$ the reference temperature and $\beta = 1/(k_{\rm B} \bar{T})$, the DPD equations read~\cite{HK92,EW95}
\begin{equation}
\label{DPD}
\left \{ \begin{array}{cl}
dq_i = & \dps \frac{p_i}{m_i} \, dt \\
dp_i = & \dps \sum_{j \not = i} -\nabla V(r_{ij}) \, dt - \gamma \chi^2(r_{ij}) 
( v_{ij} \cdot e_{ij} ) e_{ij} + \sqrt{\frac{2 \gamma}{\beta}} \chi(r_{ij}) dW_{ij} \, e_{ij} \\
\end{array} \right. 
\end{equation}
with
\[
\gamma > 0, \quad r_{ij} = |q_i-q_j|, \quad e_{ij} = \frac{q_i-q_j}{r_{ij}}, \quad v_{ij} = \frac{p_i}{m_i} - \frac{p_j}{m_j},
\]
$\chi$ a weight function (with support in $[0,r_c]$ where
$r_c$ is a cut-off radius), and where $W_{ij}$ are $1$-dimensional
independent Wiener processes such that $W_{ij}= W_{ji}$. 

Notice that, since the dissipation term depends only on the relative velocities, the dynamics are globally Galilean invariant. Besides, the total momentum is preserved. However, the total energy fluctuates, so that some refinements in the model are required. Relying on the general DPD picture, DPD with conserved energy were introduced in~\cite{AM97,Espanol97}. The idea is that the variations of the total mechanical energy
\[
H(q,p) = \frac{1}{2} p^T M p + {\cal V}(q)
\] 
through the dissipative forces are compensated by some reservoir energy variable attached to each particle. Introducing an internal energy $\epsilon_i$ for each particle, the evolution of the internal energies are constructed such that $dH(q,p) + \sum_i d\epsilon_i = 0$. 
An associated entropy $s_i = s(\epsilon_i)$ and an internal temperature can be also defined for each particle as
\[
T_i = \left ( \frac{\partial s_i}{\partial \epsilon_i} \right )^{-1}.
\]
For example, when the internal degrees of freedom are purely harmonic,
$T(\epsilon) = \epsilon/C_v$,
where $C_v$ is the specific heat at constant volume. More generally, this microscopic state law should be computed using MD or {\it ab initio} simulations.

\section{A simplified model for shock waves}
The model we consider is strongly inspired from DPD models with conserved energy~\cite{AM97,Espanol97}, so that all the properties of the usual DPD models with conserved energy can be straightforwardly transposed to this case. The derivation of the model is done as in~\cite{AM97,Espanol97}. 

The main differences here is that (i) we present the dynamics for particles of unequal masses, and (ii) do not project the dissipatives and random forces along the lines of center of the particles. The generalization to particles of unequal masses is done by considering dissipation forces depending on the relative velocities, and not on the relative momenta. This is important if mixtures composed of (say) two molecules are simulated, and each molecule is replaced by a single particle, whose mass is the total mass of the molecule. The dissipative and random forces could be projected as well to conserve angular momentum, but we restrict ourselves to the simpler and more general case when these forces are not projected, since we are only interested in Galilean invariance, and have in mind an extension to reduced models for reactive shock waves~\cite{MSS}, which do not necessarily preserve angular momentum, even if the dissipative and random forces are projected. Such a model is also closer to the Langevin picture of the previous reduced models for shock waves~\cite{Holian03,SH04}. 

We finally neglect the thermal conduction here, since the contribution to the evolution of the internal energy arising from the dissipation forces is expected to be dominant in the nonequilibrium zone near the shock front. Heat diffusion plays a role only after the relaxation towards equilibrium in the shocked zone is achieved.

The equations of motion for the system read:
\begin{equation}
\label{DPD_gen}
\left \{ \begin{array}{cl}
dq_i = & \dps \frac{p_i}{m_i} \, dt \\
dp_i = & \dps \sum_{j, \, j \not = i} -\nabla V(r_{ij}) \, dt - \gamma_{ij} \chi^2(r_{ij}) v_{ij} \, dt + \sigma_{ij} \chi(r_{ij}) dW_{ij}, \\
\end{array} \right.
\end{equation}
where $\chi$ is still a weight function (with support in $[0,r_c]$ where
$r_c$ is a cut-off radius), and $W_{ij}$ are now $d$-dimensional independents Wiener processes such that $W_{ij}=-W_{ji}$. The friction $\gamma_{ij}$ and the fluctuation magnitude $\sigma_{ij}$ will be precised below. As for DPD models with conserved energy, the dynamics is postulated in a manner such that the total energy 
\[
E(q,p,\epsilon) = H(q,p)+\sum_i \epsilon_i
\] 
is preserved. The evolution of $dH = -\sum_i d\epsilon_i$ is inferred from~(\ref{DPD_gen}) using Ttô rule (see~\cite{Espanol97} for more details).
Therefore, we consider the following dynamics:
\begin{equation}
\left \{ \begin{array}{cl}
\label{DPDsimplifie}
dq_i = & \dps \frac{p_i}{m_i} \, dt \\
dp_i = & \dps \sum_{j, \, j \not = i} -\nabla V(r_{ij}) \, dt - \gamma_{ij} \chi^2(r_{ij}) v_{ij} \, dt + \sigma_{ij} \chi(r_{ij}) dW_{ij}, \\
d\epsilon_i = & \dps \frac12 \sum_{j, \, j \not = i} \left ( \chi^2(r_{ij}) \gamma_{ij} v_{ij}^2 - \frac{\sigma_{ij}^2}{2} \left(\frac{1}{m_i}+\frac{1}{m_j}\right) \chi^2(r_{ij}) \right ) \, dt - \sigma_{ij} \, \chi(r_{ij}) v_{ij} \cdot dW_{ij}, 
\end{array} \right.
\end{equation}
with the following fluctuation-dissipation relation~\cite{AM97,Espanol97}
\begin{equation}
\label{FDR}
\sigma_{ij} = \sigma^2, \quad \gamma_{ij} = \frac{\sigma^2}{\beta_{ij}} \ \ {\rm with} \ \ \beta_{ij} = \frac{1}{2 k_{\rm B}} \left ( \frac{1}{T_i} + \frac{1}{T_j} \right ).
\end{equation}
It is then easily checked that measures of the form
\begin{equation}
\label{inv_mes}
d\rho(q,p,\epsilon) = \frac1Z {\rm e}^{-\beta H(q,p)} \exp\left(\sum_i \frac{s(\epsilon_i)}{k_{\rm B}}-\beta\epsilon_i\right) \, dq \, dp \, d\epsilon
\end{equation}
are invariant~\cite{AM97}. This measure expresses the fact that the translational degrees of freedom are distributed according to a classical Boltzmann statistics, whereas the internal energies are distributed according to some free energy statistics. 
Since the total momentum $P_0=\sum_i p_i$ and the total energy $E_0=E(q,p,\epsilon)$ are also preserved by construction, the measure
\begin{equation}
\label{inv_mes_inv}
d\rho(q,p,\epsilon) = \frac{1}{Z_{P,E}} {\rm e}^{-\beta H(q,p)} \exp\left(\sum_i \frac{s(\epsilon_i)}{k_{\rm B}}-\beta\epsilon_i\right) \delta_{E=E_0} \, \delta_{P=P_0} \, \, dq \, dp \, d\epsilon
\end{equation}
is an invariant measure. 

If the dynamics is ergodic for the measure~(\ref{inv_mes}), it holds
\[
k_{\rm B} \langle T_{\rm kin} \rangle = \beta^{-1}, \quad 
k_{\rm B} \left ( \left \langle \frac{1}{T_{\rm int}} \right \rangle \right )^{-1} = \beta^{-1},
\]
with $T_{\rm kin} = \frac{1}{k_{\rm B}dN} \sum_{i=1}^N \frac{p_i^2}{m_i}$, $\frac{1}{T_{\rm int}} = \frac{1}{N} \sum_{i=1}^N \frac{1}{T_i}$, and 
$\langle A \rangle = \int A(q,p) \, \rho(q,p,\epsilon) \, dq \, dp \, d\epsilon$. 
Notice that these relationships provide estimators for the local thermodynamic temperature $\beta^{-1}/k_{\rm B}$ through the arithmetic average kinetic temperatures, and the {\it harmonic} average internal temperatures. Let us emphasize that a straightforward arithmetic average over the internal temperatures would give wrong results (the corresponding estimator being biased). The same result holds in the case when there are invariants of the dynamics (with the invariant measure~(\ref{inv_mes_inv})) in the limit $N \to +\infty$.

\section{A deterministic version of the model}
We intend here to introduce a deterministic version of our model, which allows to bridge the gap between a previous mesoscopic deterministic model~\cite{SH04} and the DPD framework for shock waves.
The model proposed in~\cite{SH04} introduces damping forces on the
position variable directly (and not on the momentum variables as would
be expected) in order to preserve the Galilean invariance. Indeed, the damping terms in the momentum variable are considered
to be of the form $-\gamma(v_i-\bar{v}_i)$, where $\bar{v}_i$ is a local
average of the velocities around the particle, which makes
the Galilean invariance of the dissipated energy difficult to
preserve. If on the other hand the dissipation term in the momentum
variable implies only pairwise velocity differences as for DPD models,
the Galilean invariance follows immediately. The
following equations of motion then mix the deterministic equations of motion
of~\cite{SH04} and the DPD philosophy:
\[
\left \{ \begin{array}{cl}
dq_i = & \dps \frac{p_i}{m_i} \, dt \\
dp_i = & \dps \sum_{j, \, j \not = i} -\nabla V(r_{ij}) \, dt -
\gamma \frac{T^{\rm ext}_{ij}  - T_{ij}^{\rm int}}{\bar{T}} \omega(r_{ij}) v_{ij} \, dt, \\
d\epsilon_i = & \dps \frac12 \sum_{j, \, j \not = i} \gamma
\frac{T^{\rm ext}_{ij}  - T_{ij}^{\rm int}}{\bar{T}}  \omega(r_{ij})
v_{ij}^2 \, dt.
\end{array} \right. 
\]
where $T^{\rm ext}_{ij}$ is the average temperature in the kinetic
degrees of freedom of particles $i$ and $j$ (for example, $T^{\rm ext}_{ij}
= ( T^{\rm ext}_i + T^{\rm ext}_j ) /2$ with $T^{\rm ext}_i =
2 p_i^2 / k_{\rm B}d m_i$ the kinetic
temperature associated with particle $i$)
and $T_{ij}^{\rm int}$ is the average internal temperatures of particles
$i$ and $j$ (for example, $T^{\rm int}_{ij}
= ( T^{\rm int}_i + T^{\rm int}_j ) /2$). The function $\omega$ is still
a weighting function, and $\gamma$ determines the strength of the
coupling. 

Notice that the dissipation term is in fact a dissipation term only when 
$T^{\rm ext}_{ij} > T_{ij}^{\rm int}$, and an anti-dissipation term
otherwise. This ensures that the internal and
external (kinetic thus potential terms) energies equilibriate in all cases. 
However, the thermodynamic properties of such a model are less
clear to state than for the previous stochastic model, and so, we stick
to the model~(\ref{DPDsimplifie}).

\section{Numerical discretization}
\label{section_scheme}

We use splitting formulas inspired from~\cite{shardlow03,SdFEC06}. Recall that the integration of the equation of motion~(\ref{DPDsimplifie}) is not straightforward since the dissipation terms depend on the relative velocities.

We decompose~(\ref{DPDsimplifie}) into elementary SDEs, and denote by $\phi_{\Delta t}$ the (stochastic) flow map for a time $\Delta t$. The elementary SDEs are the usual deterministic Newton part
and the dissipation part, which read respectively
\[
\left \{ \begin{array}{cl}
dq = & M^{-1} p \, dt, \\
dp = & -\nabla V(q) \, dt
\end{array} \right.
\quad {\rm and} \quad
\forall i < j, \quad \left \{ \begin{array}{cl}
dp_i = & -\gamma_{ij} \chi^2(r_{ij}) v_{ij} \, dt + \sigma \chi(r_{ij}) \, dW_{ij}, \\
dp_j = & -dp_i, \\
d\epsilon_i = & -\frac12 d\left ( \frac{p_i^2}{2m_i} + \frac{p_j^2}{2m_j}\right ), \\
d\epsilon_j = & d\epsilon_i.\\
\end{array} \right.
\]
Denoting by $\phi_{{\rm Newton}, \Delta t}$ and $\phi^{i,j}_{{\rm diss}, \Delta t}$ ($1 \leq i < j \leq N$) the associated stochastic flow maps, an approximation of $\phi_{\Delta t}$ of order 1 is $\phi_{\Delta t} \simeq
\phi^{1,2}_{{\rm diss}, \Delta t} \circ \dots \circ 
\phi^{N-1,N}_{{\rm diss}, \Delta t} \circ
\phi_{{\rm Newton}, \Delta t}$.
Notice that in practice it is difficult to resort to a second order accuracy sheme through a Strang splitting
since this requires to keep track of the order the variables were updated in the first part of the time step, which is cumbersome when using Verlet lists.

The Newton flow $\phi_{{\rm Newton},\Delta t}$ is approximated using a Velocity-Verlet scheme.
For an approximation $\Phi^{i,j}_{{\rm diss},\Delta t}$ ($i < j$) of 
the dissipation part, we first update the velocities at fixed internal temperatures
using a Verlet-like algorithm as proposed in~\cite{shardlow03}:
\[
\left \{ \begin{array}{cl}
p_i^{n+1/2} = \dps p_i^n - \frac12 \gamma_{ij} \chi^2(r_{ij}) v_{ij}^n + \frac12 \sigma \sqrt{\Delta t} \chi(r_{ij}) \, U_{ij}^n, \\
p_j^{n+1/2} = \dps p_j^n + \frac12 \gamma_{ij} \chi^2(r_{ij}) v_{ij}^n - \frac12 \sigma \sqrt{\Delta t} \chi(r_{ij}) \, U_{ij}^n, \\
p_i^{n+1} = \dps p_i^{n+1/2} - \frac12 \gamma_{ij} \chi^2(r_{ij}) v_{ij}^{n+1} + \frac12 \sigma \sqrt{\Delta t} \chi(r_{ij}) \, U_{ij}^n, \\
p_j^{n+1} = \dps p_j^{n+1/2} + \frac12 \gamma_{ij} \chi^2(r_{ij}) v_{ij}^{n+1} - \frac12 \sigma \sqrt{\Delta t} \chi(r_{ij}) \, U_{ij}^n, \\
\end{array} \right.
\]
where $(U_{ij}^n)_{1\leq i<j\leq N, n \geq 0}$ are independently and identically distributed standard gaussian random variables.
Notice that the third and fourth updates can be rewritten in an explicit form~\cite{shardlow03}. The energy is then updated as
\[
\epsilon_i^{n+1} - \epsilon_i^n = \epsilon_j^{n+1} - \epsilon_j^n =
\frac12 \left ( \frac{(p_i^{n+1})^2}{2m_i} + \frac{(p_j^{n+1})^2}{2m_j}- \frac{(p_i^n)^2}{2m_i} - \frac{(p_j^n)^2}{2m_j} \right).
\]
so that the total energy is indeed conserved by this step. Of course, this integration scheme could be refined, especially the dissipation part. We however tested several refinements and/or higher orders integrations in some model cases, but found no noticeable differences in the results.

\section{Application to shock waves}
\label{res_num}

Some numerical simulations of DPD models with conserved energy where proposed in~\cite{RE98,AM99}, but were concerned only with the computation of thermal conductivities. The corresponding nonequilibrium states were stabilized using steady temperature gradients. The dissipation terms in the DPDE equations of motions were discarded, and only the diffusive part was retained. 
We present in this section profiles obtained from simulations of shock waves, for which the diffusive part of the dynamics can be discarded, but the dissipative part is of paramount importance to reproduce qualitative and quantitative features of all-atom shock waves. This situation is somehow complementary to the cases studied in~\cite{RE98,AM99}, and, to our knowledge, was never considered before for some physical application. 

We consider the crystalline polymer (PVDF) system of~\cite{SH04}, the corresponding reduced system being modeled by a two-dimensional (2D) triangular lattice of mesoparticles. Results for the all-atom model can also be found in~\cite{SH04}.

The effective interaction potential between mesoparticles is a pairwise Rydberg potential of the form~\cite{SH04}
\[
V(r) = V_{\rm R}\left(\lambda\left(\frac{r}{r_0}-1\right)\right) \quad {\rm with} \ \ V_{\rm R}(d) = - \epsilon \, (1 + d + \alpha d^3) \, {\rm e}^{-d}.
\]
The parameters given by~\cite{SH04} were fitted to reproduce the stress
in an uniaxial compression: $\lambda = 7.90$, $\alpha = 0.185$, $r_0 =
5.07$~\AA, $\epsilon = 1.612 \times 10^{-20}$~J, $m = 64.03 \times 10^{-3}$~kg/mol. We also choose a cut-off radius $R_{\rm cut} = 15$~\AA \ for the pairwise interactions.
The microscopic state law is obtained by assuming that $C_v$ is independent of the temperature: $\epsilon = C_v T$,
with here $C_v = 16 \ k_{\rm B}$ since we represent a three-dimensional molecule formed of 6 atoms by a 2D mesoparticle. In general, the heat capacity is a function of the temperature $C_v = C_v(T)$, and should be parametrized by equilibrium simulations. 

We use the simple weight function $\chi(r) = ( 1 - r/R_{\rm cut})^2$ if $r \geq R_{\rm cut}$, $\chi(r) = 0$ otherwise,
the cut-off radius $R_{\rm cut}$ being the same as the one used for the potential. Of course, many other weight functions could be used. We also set $\gamma = 1.5 \times 10^{-14}$~kg/s and $\Delta t = 10^{-14}$~s. In these preliminary tests of the model, the parameter $\gamma$ was varied to obtain a good agreement with the all-atom results. However, it is expected that $\gamma$ is linked to some physical quantity, such as the decay rate of the relative velocities autocorrelation in an all-atom simulation, and could therefore be estimated using some preliminary small equilibrium simulations.

We first prepare an initial state according to the invariant measure~(\ref{inv_mes}). To this end, we sample independently the internal energies according to the measure $Z_\epsilon^{-1} \exp(-\beta \epsilon + s(\epsilon)/k_{\rm B}) = Z_\epsilon^{-1} \epsilon^{C_v/k_{\rm B}} \exp(-\beta \epsilon)$, and the initial configuration in phase-space by thermalizing a lattice initially at rest, using a Langevin dynamics. 
In this study, the initial temperature is $T_0 = 300$~K, and the edge of the triangles in the triangular lattice is $a = 5.13$~\AA.

\begin{figure}
\begin{center}
\includegraphics[angle=270,width=144mm]{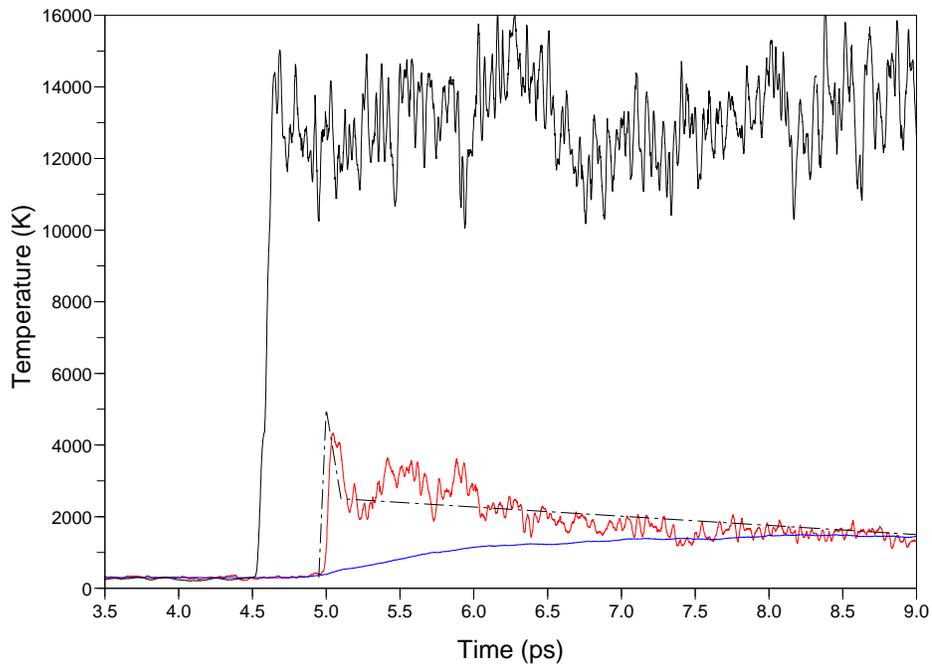}
\caption{ \label{figure_choc_inerte} (color online) 
Temporal evolution of the temperature of a thin slab of material as the shock runs trough it~: mean kinetic
  temperature $\hat{T}_{\rm kin}$ in the direction of the shock
  (intermediate curve, red), mean internal temperature $\hat{T}_{\rm int}$ (lower curve, blue). The corresponding results when the coupling with the
  internal degress of freedom is turned off are also shown (upper curve,
  black), and a cartoon representation of the all-atom result from~\cite{SH04} for the kinetic
  temperature $\hat{T}_{\rm kin}$ is also plotted (dark dashed line).}
\end{center}
\end{figure}

We then produce a shock using a piston at velocity $u_{\rm p} = 3000$~m/s.
Figure~\ref{figure_choc_inerte} presents the relaxation behind the shock
front for the 2D triangular lattice of mesoparticles subjected to the
dynamics~(\ref{DPDsimplifie}).
The results are in good agreement with the all-atom results of~\cite{SH04}. In particular, the final temperature is very close to the all-atom value (whereas it is of course greatly overestimated by the mesoscopic dynamics without coupling), and the time required for the internal temperatures and kinetic temperatures to equilibriate is almost the time needed in all-atom studies. 


\section{Conclusion and prospects}

We have proposed a simplified DPDE dynamics~\cite{AM97,Espanol97}, and shown its ability to simulate shock waves in a simple model case. 
This method allows a computational gain of one order of magnitude in space in the particular example considered, since 18 degrees of freedom are replaced by 2 degrees of freedom (assuming that the computational cost scales linearly with respect to the number of particles, which is the case here since we resort to short-ranged potentials).  
Besides, the time steps used can be taken larger as the time steps required by straightforward MD simulations since do not need to resolve the high-frequency atomic vibrations.

Further quantitative agreement could be obtained by resorting to a less simplistic microscopic state law $T = T(\epsilon)$, and parametrizing more systematically the dynamics. In particular, the friction coefficient $\gamma$ could be obtained from equilibrium all-atom simulations. Memory effects could also be introduced by resorting to history-dependent friction
terms (in analogy with the so-called generalized Langevin equations).

More importantly, this study opens the way to a reduced model of
detonation waves. For detonation waves, chemical reactions have to be
treated explicitely and accurately in order to obtain the right
velocities. Work is in progress to incorporate the corresponding
chemical processes in the current DPDE framework~\cite{MSS}.

\section*{Acknowledgments}
I warmly thank Alejandro Strachan and Brad Holian for their helpful and stimulating comments and remarks, and Jean-Bernard Maillet and Laurent Soulard for their help with the simulation code S-TAMP. This work was supported by the ACI grant "Simulation moléculaire" from the french Ministry of Research.

%
%

\end{document}